\documentclass[preprint,authoryear,12pt]{elsarticle}
\usepackage{graphicx}
\usepackage{a4wide}
\usepackage{epstopdf}
\usepackage{booktabs}
\usepackage{natbib} 
\usepackage{multirow}
\usepackage{amssymb}
 \usepackage{amsthm}
 \usepackage{amsmath}

\journal{
Physica A
}

\begin{document}

\begin{frontmatter}

\title{Gold, currencies and market efficiency}

\author[utia]{Ladislav Kristoufek} \ead{kristouf@utia.cas.cz}
\author[utia]{Miloslav Vosvrda} \ead{vosvrda@utia.cas.cz}

\address[utia]{Institute of Information Theory and Automation, Academy of Sciences of the Czech Republic, Pod Vodarenskou Vezi 4, 182 08, Prague, Czech Republic, EU} 

\begin{abstract}
Gold and currency markets form a unique pair with specific interactions and dynamics. We focus on the efficiency ranking of gold markets with respect to the currency of purchase. By utilizing the Efficiency Index (EI) based on fractal dimension, approximate entropy and long-term memory on a wide portfolio of 142 gold price series for different currencies, we construct the efficiency ranking based on the extended EI methodology we provide. Rather unexpected results are uncovered as the gold prices in major currencies lay among the least efficient ones whereas very minor currencies are among the most efficient ones. We argue that such counterintuitive results can be partly attributed to a unique period of examination (2011-2014) characteristic by quantitative easing and rather unorthodox monetary policies together with the investigated illegal collusion of major foreign exchange market participants, as well as some other factors discussed in some detail.
\end{abstract}


\end{frontmatter}

\newpage

\section{Introduction}

For decades, the efficient market hypothesis (EMH) has been a building block of financial economics. In his fundamental paper, \cite{Fama1970} summarizes the then-current empirical findings following the theoretical papers of \cite{Fama1965} and \cite{Samuelson1965}. \cite{Fama1991} then recalls various issues of the hypothesis and reviews the newer literature on the topic. The capital market efficiency is standardly parallelized with the informational efficiency so that the markets are efficient as long as all the available information is fully reflected into market prices \citep{Fama1970}. Depending on the level of information availability, the EMH is usually separated into three forms -- weak (historical prices), semi-strong (public information), and strong (all information, even private) \citep{Fama1991}. The theory has been challenged on both theoretical \citep{Malkiel2003} and empirical \citep{Cont2001} grounds regularly, yet still it remains a popular and fruitful topic of financial research.

The empirical testing of capital markets efficiency has a long history across various assets. The already-mentioned review study of \cite{Fama1970} focuses mainly on stock markets. In commodity markets, \cite{Roll1972} and \cite{Danthine1977} are among the first ones to study their efficiency arriving at contradicting results. In the same timeline, foreign exchange rates are investigated as well \citep{Frenkel1976,Cornell1978}. The termination of the Bretton Woods system in 1971 made the detachment of gold and currency prices interesting for research of the separate phenomena \citep{Booth1979}. Nonetheless, the two still remain tightly connected. \cite{Koutsoyiannis1983} focuses on the efficiency of gold prices and argues that the market efficiency cannot be refuted. Nevertheless, the author finds a tight connection between gold prices and the strength of the US dollar as well as the inflation, interest rates and a general state of the US economy. The gold prices and foreign exchange rates are thus found to be firmly interconnected, which is supported by another early study of \cite{Ho1985}. \cite{Frank1989} further suggest that simple linear testing of the gold (and silver) market efficiency need not be sufficient.

The efficiency studies of foreign exchange rates are quite unique compared to the mentioned stocks and commodities as the foreign exchange rates pricing has solid macroeconomic foundations such as the balance of payment theory, the purchasing power parity, the interest rate parity, the Fisher effect and others \citep{DunnJr.2004,Levi2005,Feenstra2008}. These theories lead to different ways of efficiency treatment and testing.

\cite{Charles2012} examine the return predictability of major foreign exchange rates between 1975 and 2009. Using various tests, the authors show that the exchange rates are unpredictable most of the time. Short-term inefficiencies are attributed to major events such as coordinated central bank interventions and financial crises. The crises perspective is further studied by \cite{Ahmad2012} who focus on the Asia-Pacific region. They argue that the 1997-1998 Asian crisis was more disturbing compared to the 2008-2009 global financial crisis. In addition, the floating currency markets are found to be more resilient than the countries with managed currencies. \cite{Al-Khazali2012} further examine the Asia-Pacific region using the random walk and martingale definitions of the market efficiency. Out of 8 studied currencies, only three (Australian dollar, Korean won and Malaysian ringgit) are found to be efficient while the other exchange rates offer profitable trading opportunities.

\cite{Olmo2011} review the literature on the foreign exchange rate efficiency testing based on the uncovered interest rate parity. They suggest that the rejection of efficiency in this area of research may be due to significant differences in volatilities of the logarithmic changes of exchange rates and the forward premium, in addition to conditional heteroskedasticity of the data. The authors introduce a set of profitability-based tests of market efficiency based on the uncovered interest rate parity and they show that the foreign exchange rates are much closer to market efficiency than usually claimed. \cite{Chen2013} inspect whether interest rates structure (yield curve) can be used for foreign exchange rate forecasting. They show that it is the case on time horizons between one month and two years. They also argue that these results can help explaining the uncovered interest rate parity puzzle by relating currency risk premium to inflation and business cycle risks. \cite{Bianco2012} further discuss the potential of using economic fundaments for foreign exchange rates forecasting. Their fundamentals-based econometric model for weekly euro-dollar rates is shown to beat the random walk model for time horizons between one week and one month. \cite{Engel2015} construct factors from exchange rates and they use their idiosyncratic deviations for forecasting. Combining these with the Taylor rule, and monetary and purchasing power parity models, they improve the forecasting power of the model compared to the random walk benchmark for the periods between 1999 and 2007 but not for earlier periods down to 1987.

\cite{Chaboud2014} inspect the effect of algorithmic trading on efficiency of the foreign exchange markets in the high-frequency domain. They show that algorithmic trading improves market efficiency in two aspects -- triangular arbitrage opportunities and autocorrelation of high-frequency returns. On the contrary, they argue that this may impose higher adverse selection costs on slower traders.

Studies of the foreign exchange rates efficiency, in the same way as of the other assets, primarily focus on testing whether a given currency or a set of currencies may or may not be considered efficient. To reflect this point, \cite{Kristoufek2013} introduce the Efficiency Index (EI) which can be used to rank assets according to their efficiency. In addition, the index is very flexible and it can incorporate various measures of the market efficiency. In the original study, \cite{Kristoufek2013} study 41 stock indices and find the Japanese NIKKEI to be the most efficient one. From a geographic perspective, the most efficient indices are localized in Europe and the least efficient ones in Asia and Latin America. \cite{Kristoufek2014} further focus on the index specification and show that approximate entropy adds a significant informative value to the index. \cite{Kristoufek2014a} then study efficiency across various commodity futures and uncover that energy commodities are the most efficient ones whereas the livestock commodities such as cattle and hogs are the least efficient ones. Here we focus on efficiency ranking of the gold market with respect to a currency used for the purchase, and we also contribute to the discussion on statistical properties of the Efficiency Index. 

\section{Methods}

Coming back to the roots of the efficient market hypothesis in 1965, the treatment has been split into two main branches -- based on the random random walk hypothesis \citep{Fama1965} and following the martingale specification \citep{Samuelson1965}. We follow the latter approach as it is less restrictive and it assumes the returns of the efficient market to be only serially uncorrelated and with finite variance. This straightforward treatment enables us to use various measures of market efficiency and use them to construct the Efficiency Index, which allows to rank financial assets according to their efficiency. In this section, we briefly describe the Efficiency Index, its components and its statistical treatment. Introducing a procedure to assess statistical features of the Efficiency Index is an important and novel contribution to this line of research.

\subsection{Capital market efficiency measure}

\cite{Kristoufek2013,Kristoufek2014a,Kristoufek2014} define the Efficiency Index (EI) as
\begin{equation}
\label{eq:EI}
EI=\sqrt{{\sum_{i=1}^n{\left(\frac{\widehat{M_i}-M_i^{\ast}}{R_i}\right)^2}}},
\end{equation}
where $M_i$ is the $i$th measure of efficiency, $\widehat{M_i}$ is an estimate of the $i$th measure, $M_i^{\ast}$ is an expected value of the $i$th measure for the efficient market and $R_i$ is a range of the $i$th measure. EI is thus a distance from the efficient market situation. The index can include various efficiency measures but these need to be bounded, which turns out to be rather restrictive. We utilize three efficiency measures, which meet such criterion and which are frequently used in market efficiency studies \citep{Cajueiro2004a,Cajueiro2005,DiMatteo2005,DiMatteo2007,Zunino2010,Zunino2011,Ortiz-Cruz2012} -- Hurst exponent $H$ with an expected value of 0.5 for the efficient market ($M_H^{\ast}=0.5$), fractal dimension $D$ with an expected value of 1.5 ($M_D^{\ast}=1.5$), and the approximate entropy with an expected value of 1 ($M_{AE}^{\ast}=1$). As discussed later in this section, Hurst exponent and fractal dimension share their range for stationary processes whereas approximate entropy does not. For this point, we need to rescale the approximate entropy part of the Efficiency Index so that we have $R_{AE}=2$ and $R_D=R_H=1$.

\subsection{Long-range dependence and its estimators}

Long-range dependent series can be formally described as the ones with a power-law decaying autocorrelation function (in time domain) and/or a divergent at origin spectrum (in frequency domain). Specifically, the autocorrelation function $\rho(k)$ with time lag $k$ of a long-range dependent process decays as $\rho(k) \propto k^{2H-2}$ for $k\rightarrow +\infty$, and spectrum $f(\lambda)$ with frequency $\lambda$ scales as $f(\lambda)\propto \lambda^{1-2H}$ for $\lambda \rightarrow 0+$ \citep{Geweke1983,Beran1994,Robinson1995a}. The characteristic parameter $H$ is Hurst exponent which has several interesting values and intervals of existence. For $H<0.5$, the processes are anti-persistent and switch their sign frequently compared to an uncorrelated process. For $H=0.5$, the processes are not long-range dependent, and for $H>0.5$, the processes are persistent. The last group of processes can be further categorized according to stationarity and (non)existence of variance. For stationary processes, it holds that $H<1$. For the purposes of the Efficiency Index construction, it is important that for an efficient market, we have $H=0.5$, as well as is the fact that the index is bounded for stationary processes. Out of plethora of Hurst exponent estimators \citep{Beran1994,Taqqu1995,Taqqu1996,Robinson1995a,Geweke1983,DiMatteo2003,DiMatteo2007,Barunik2010,Teverovsky1999}, we choose the local Whittle estimator and the GPH estimator as they are suitable for short time series with possible weak short-term memory, and they are consistent and asymptotically normal \citep{Geweke1983,Beran1994,Robinson1995a,Taqqu1995,Taqqu1996,Phillips2004}.

\subsection{Fractal dimension}

Long-range dependence can be seen as a global characteristic of a time series. Contrary to this view, fractal dimension $D$ can be interpreted as a measure of local memory of the series since it captures roughness of the series \citep{Kristoufek2013}. Fractal dimension ranges between $1<D\le 2$ for univariate series and this range is separated by the value of $D=1.5$ for uncorrelated processes, which represents the efficient markets value. Low fractal dimension signifies lower roughness and thus local persistence. Reversely, high fractal dimension characterizes rougher series and thus locally negatively correlated. Fractal dimension is thus well defined for an efficient market and it is bounded for univariate series, which makes it a perfect candidate to be included into the Efficiency Index. Specifically, we utilize two estimators of fractal dimensions which share desirable statistical properties for short time series -- Hall-Wood and Genton estimators \citep{Gneiting2004,Gneiting2010}.

\subsection{Approximate entropy}

Entropy is considered as a measure of complexity. High entropy suggests little or no information in the system and thus high uncertainty whereas low entropy is characteristic for deterministic systems \citep{Pincus2004}. From the efficiency perspective, systems with maximum entropy can be seen as efficient as these are serially uncorrelated. The lower the entropy level, the less efficient the market is. For the construction of the Efficiency Index, we utilize the approximate entropy which is bounded and thus well suited for the index \citep{Pincus1991}.

\subsection{Statistical inference}
\label{sec}

The original Efficiency Index \citep{Kristoufek2013} is a point estimate of the true index value. This poses problems when discussing the results and their statistical validity. We tackle this issue by introducing a new approach to estimating EI which stems in the following steps:
\begin{enumerate}
\item Obtain the estimated components $\widehat{M}_i$ of the Efficiency Index according to Eq. \ref{eq:EI}.
\item Shuffle the underlying return series.
\item Estimate the components of the Efficiency Index for the shuffled series, and label these as $\widehat{M}_{i,shuffle}$.
\item Use $\widehat{M}_{i,shuffle}$ in place of $M^{\ast}_i$ in Eq. \ref{eq:EI}.
\item Obtain $\widehat{EI}$ based on the previous steps.
\item Repeat $N$ times.
\item Obtain necessary statistics based on these $N$ estimates.
\end{enumerate}
This way, we obtain an estimate of the Efficiency Index which controls for the potential finite sample bias and the influence of distributional properties of the analyzed series. For purposes of our study, we set $N=100$.

\section{Results and discussion}


We study the efficiency ranking of the gold\footnote{Gold is selected as a num\'{e}raire due to its historical reputation as a safe haven as well as its reserve status and a relative long-term price stability.} prices quoted in different currencies. The portfolio of study comprises 142 worldwide currencies, which are described in Table \ref{tab1}. The dataset has been obtained from oanda.com, which provides a large set of FX pairs as well as gold (and other precious metals) prices in various currencies. The covered period ranges between 1.1.2011 and 30.11.2014, which totals 1430 observations for each of the 142 analyzed currencies\footnote{We prefer a width of the portfolio to its depth to be able to compare as many currencies as possible.}. These currencies cover almost all available and traded fiat currencies in addition to Bitcoin, the most popular and used cryptocurrency.

For the efficiency ranking, we use the Efficiency Index (Eq. \ref{eq:EI}) with adjustments described in Sec. \ref{sec}. Specifically, we utilize two measures of long-range dependence -- the local Whittle estimator and the GPH estimator --, two measures of fractal dimension -- the Hall-Wood estimator and the Genton estimator -- and the approximate entropy as proposed by \cite{Pincus2004}. In the procedures, we follow the standard procedure of using the logarithmic returns for the Hurst exponent and approximate entropy estimators, and logarithmic prices for the fractal dimension estimation. Using 100 repetitions (shuffling), we obtain the estimated Efficiency Index as a median value with a corresponding standard error for more information about precision of the estimate.

The resulting ranking of gold prices with respect to the used currency is presented in Table \ref{tab2}. The ranking is rather unexpected or even surprising. Practically all of the most liquid currencies -- the US dollar, the British pound, the Australian dollar, the New Zealand dollar, the Japanese yen, the Euro, the South Korean won, the Norwegian krone -- are among the least efficient gold markets (the least efficient third of the sample). Among these, also the Bitcoin currency lays at the very bottom of the ranking. On the other side of the ranking, the Top 5 is formed by the the Liberian dollar, the Seychellois rupee, the Maldivian rufiyaa, the Comorian franc, and the Somali shilling. The differences between levels of EI are stunning as the most efficient markets share the index between 0.1 and 0.2 whereas the least efficient ones jump close to 0.4. Such divergence is further accentuated by very low standard errors of the estimates usually below 0.01 (medians and standard errors are reported in Table \ref{tab2}). 

\begin{figure}[htbp]
\center
\begin{tabular}{c}
\includegraphics[width=6.5in]{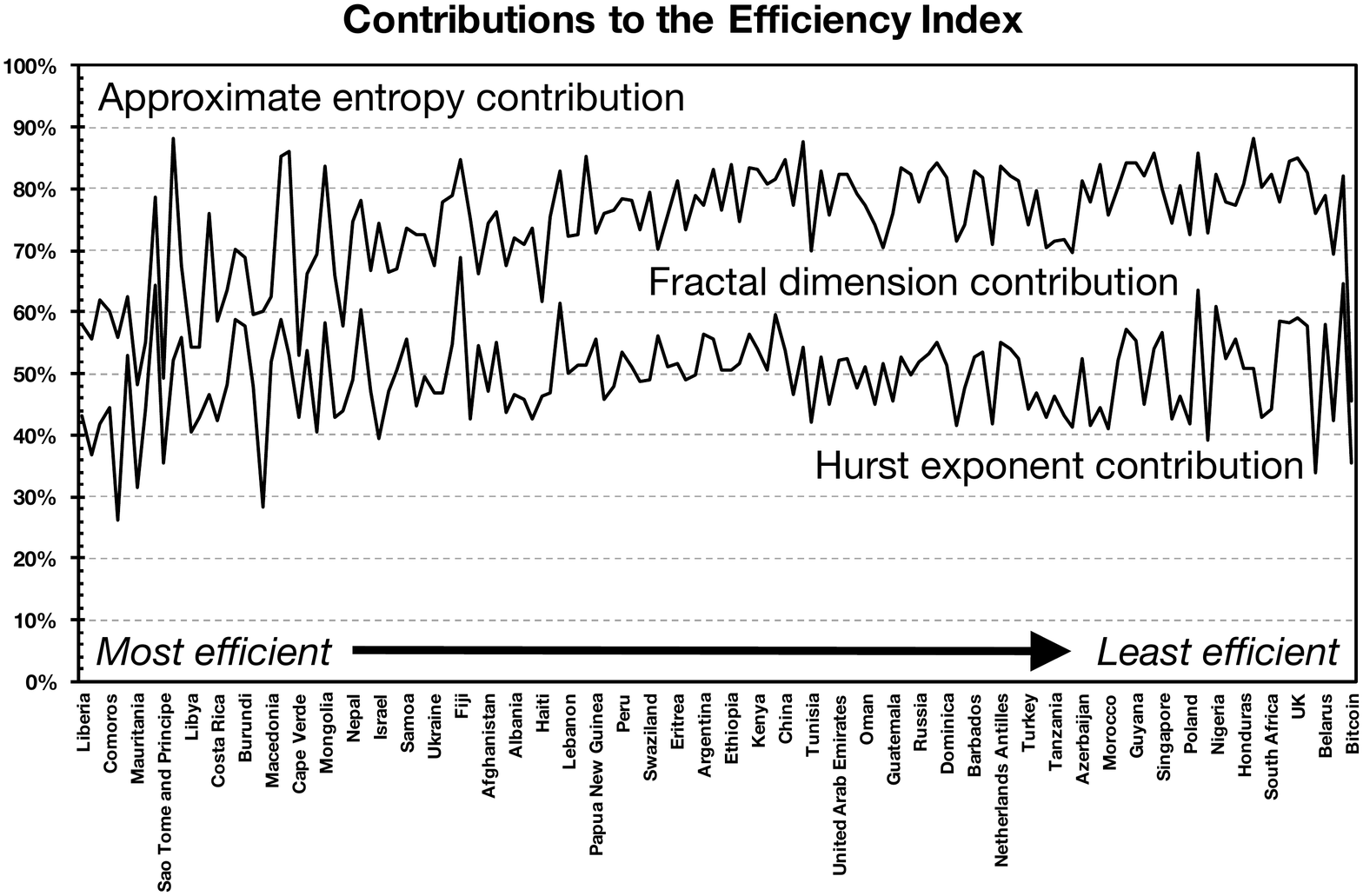}\\
\end{tabular}
\caption{\footnotesize\textbf{Contributions to the Efficiency Index.} Contributions of the three factors -- Hurst exponent, fractal dimension, and approximate entropy -- are illustrated here in percentage ($y$-axis). The currencies are ranked from the most efficient ones (\textit{from the left}) to the least efficient ones (\textit{to the right}). Note that not all the labels are visible on the $x$-axis due to the high number of analyzed currencies. Nonetheless, the values and contributions are present for all the currencies. The contributions are stacked. The bottom part represents the Hurst exponent contribution, the middle part illustrates the fractal dimension contribution, and the upper part shows the approximate entropy contribution. Hurst exponent plays an important role for all currencies and its contribution around 50\% is quite stable across all currencies. For the most efficient ones, the contributions is slightly lower. The fractal dimension contribution is quite small for the most efficient markets and its value increases for the less efficient markets. The reverse is true for the approximate entropy contribution. \label{EI}}
\end{figure}

To further investigate the contribution of the three different parts of the Efficiency Index, i.e. Hurst exponent, fractal dimension, and approximate entropy, we present Fig. \ref{EI}. We observe that overall Hurst exponent is the biggest contributor to the index. However, the strength of contribution varies with the efficiency ranking. For the most efficient currencies, Hurst exponent and approximate entropy play a similar role in the index. The influence of the latter declines with a decreasing efficiency, and vice versa for the former. The role of fractal dimension is also efficiency dependent. For the most efficient markets, it forms only a small fraction of the index but its role slightly increases for the less efficient currencies. For approximately the lower two thirds of the ranking, the contributions are rather stable with Hurst exponent at around 50\%, and fractal dimension and approximate entropy each at around 25\%. All three components of the Efficiency Index thus form its important parts. But apart from the most efficient markets, the long-term memory plays a prominent role. The gold prices in various currencies exhibit a persistent behavior with long-term trends even from the global perspective. Such stable results accentuate the advantages of using the adjusted methodology proposed here.

Returning back to the overall results which can be labelled as unexpected ones (contrary to the quite expected results found for the stock markets \citep{Kristoufek2013,Kristoufek2014} and other commodities \citep{Kristoufek2014a}), we highlight the specific connection between the gold market and the currency markets and further discuss potential causes.

The analyzed period of 2011 and 2014 covers very unorthodox times with regards to monetary policies of the developed world as reactions to the Global financial crisis, the Eurozone crisis, the Greek crisis and connected phenomena. Various waves of the quantitative easing (QE) in the USA and the UK, together with parallel actions of the European Central Bank eventually leading to the quantitative easing as well, have formed an enormous pressure on the relevant currencies and their depreciation. The first two waves of QE in the USA pushed the gold prices upwards as these rallied till the end of 2011. The last wave of the USA QE, which was much weaker than the previous two had no significant effect on the USD gold prices. The connection between the currency depreciation and the consequent gold price (in the given currency) boosting, together with a long-term effect of QE known in advance forms a perfect environment for inefficiency of the gold market. This is well in hand with most of the currencies the central banks of which participated in QE or other forms of practical money-printing being among the least efficient markets. It also further puts forward the gold's speculative asset status during the QE periods.

Such reasoning is further supported by gold being used as a hedge against inflation \citep{Narayan2010}. During the initial stages of QE, there was a serious concern about uncontrolled inflation as a reaction to the virtual money-printing. As investors were hedging against expected inflation by purchasing gold, its price was pushed further up. In time, the concern slowly vanished as there were no signs of dangerous inflation pressures. Nonetheless, the predictability and inefficiency of the gold markets under QE currencies have come out as the final effect.

The following complementary explanation of the ranking structure can be quite counterintuitive in the efficient market logic. The fact that a central bank or a central authority of a country is transparent and holds up to its word can actually lead to market inefficiency. Consider a central bank announcing a new wave of QE. If the central bank is trustworthy, the investors will start behaving accordingly and maximize their profit by acting upon it. However, the QE process is a gradual one and it thus does not affect the market instantly but in steps. Putting these factors together leads us to a quite well predictable market behavior with relatively low risk assuming the authority holds up to its promises. From the other side, the authorities which are not too trustworthy are prone to change the announced policy repeatedly so that the shocks to the currency market are unpredictable. Such unpredictability leads to higher efficiency.

Additionally, the current situation at the foreign exchange markets has uncovered another potential source of inefficiency in the most traded currencies. In 2013, Bloomberg News reported that global regulators had started investigation of major banks in the foreign exchange markets for front-running orders and colluding to rig the foreign exchange rate benchmarks \citep{BoE2014}. This affair is now referred to as the ``forex probe'' and it has been claimed that the exchange rates manipulation had been realized for about a decade. Such collusion goes majorly against the notion of market efficiency and it provides a firm ground to the reported results of currency efficiency ranking.

To summarize, the combination of gold prices and currencies forms a very interesting and unique structure, the dynamics of which is much different compared to other assets such as stock or commodity markets. We have shown that the least efficient gold prices are mostly the ones quoted in major currencies such the US dollar, the Euro, and the British pound. On the other side of the spectrum, the most efficient gold prices are the ones quoted in smaller and less traded currencies. From the practitioners' perspective, we have two possibilities of utilizing the results. We can either speculate on gold prices in the major currencies, or we can hedge gold prices using the minor currencies to obtain stable and efficient market position. Only the time will tell whether the ``forex probe'' scandal and its resolution as well as the end of the quantitative easing(s) will bring major currencies closer to efficiency.


\section*{Acknowledgements}
 
The research leading to these results has received funding from the European Union's Seventh Framework Programme (FP7/2007-2013) under grant agreement No. FP7-SSH-612955 (FinMaP) and the Czech Science Foundation project No. P402/12/G097 ``DYME -- Dynamic Models in Economics''.

\bibliography{EI_FX}
\bibliographystyle{chicago}

\newpage

\begin{table}[htbp]
\centering
\caption{Analyzed currencies}
\label{tab1}
\scriptsize
\begin{tabular}{cc|cc|cc}
\toprule \toprule
Currency name&Code&Currency name&Code&Currency name&Code\\
\midrule \midrule
Afghan afghani&AFN&Ghana cedi&GHS&Pakistani rupee&PKR\\
Albanian lek&ALL&Gibraltar pound&GIP&Panamanian balboa&PAB\\
Algerian dinar&DZD&Guatemalan quetzal&GTQ&Papua New Guinean kina&PGK\\
Angolan kwanza&AOA&Guyanese dollar&GYD&Paraguayan guarani&PYG\\
Argentine peso&ARS&Haitian gourde&HTG&Peruvian nuevo sol&PEN\\
Armenian dram&AMD&Honduran lempira&HNL&Philippine peso&PHP\\
Aruban florin&AWG&Hong Kong dollar&HKD&Polish zloty&PLN\\
Australian dollar&AUD&Hungarian forint&HUF&Qatari riyal&QAR\\
Azerbaijani manat&AZN&Icelandic krona&ISK&Romanian leu&RON\\
Bahamian dollar&BSD&Indian rupee&INR&Russian rubble&RUB\\
Bahraini dinar&BHD&Indonesian rupee&IDR&Rwandan franc&RWF\\
Bangladeshi taka&BDT&Iraqi dinar&IQD&Saint Helena pound&SHP\\
Barbadian dollar&BBD&Israeli new shekel&ILS&Samoan tala&WST\\
Belarusian ruble&BYR&Jamaican dollar&JMD&Sao Tome and Principe dobra&STD\\
Belize dollar&BZD&Japanese yen&JPY&Saudi riyal&SAR\\
Bitcoin&BTC&Jordanian dinar&JOD&Serbian dinar&RSD\\
Botswana pula&BWP&Kazakhstani tenge&KZT&Seychellois rupee&SCR\\
Brazilian real&BRL&Kenyan shilling&KES&Sierra Leonean leone&SLL\\
British pound&GBP&Kuwaiti dinar&KWD&Singapore dollar&SGD\\
Brunei dollar&BND&Kyrgyzstani som&KGS&Solomon Islands dollar&SBD\\
Bulgarian lev&BGN&Lao kip&LAK&Somali shilling&SOS\\
Burundian franc&BIF&Lebanese pound&LBP&South African rand&ZAR\\
Cambodian riel&KHR&Lesotho loti&LSL&South Korean won&KRW\\
Canadian dollar&CAD&Liberian dollar&LRD&Sri Lanka rupee&LKR\\
Cape Verdean escudo&CVE&Libyan dinar&LYD&Surinamese dollar&SRD\\
Cayman Islands dollar&KYD&Lithuanian litas&LTL&Swazi lilangeni&SZL\\
CFP franc&XPF&Macanese pataca&MOP&Swedish krona&SEK\\
Chilean peso&CLP&Macedonia denar&MKD&Swiss franc&CHF\\
Chinese yuan&CNY&Malagasy ariary&MGA&Syrian pound&SYP\\
Colombian peso&COP&Malaysian riggit&MYR&Tajikistani somoni&TJS\\
Comorian franc&KMF&Maldivian rufiyaa&MVR&Tanzanian shilling&TZS\\
Congolese franc&CDF&Mauritanian ouguiya&MRO&Thai baht&THB\\
Costa Rican colon&CRC&Mauritian rupee&MUR&Tongan pa'anga&TOP\\
Croatian kuna&HRK&Mexican peso&MXN&Trinidad and Tobago dollar&TTD\\
Cuban convertible peso&CUC&Moldovan leu&MDL&Tunisian dinar&TND\\
Czech koruna&CZK&Mongolian togrog&MNT&Turkish lira&TRY\\
Danish krone&DKK&Moroccan dirham&MAD&Turkmenistan manat&TMT\\
Djiboutian franc&DJF&Mozambican metical&MZN&Ugandan shilling&UGX\\
Dominican peso&DOP&Namibian dollar&NAD&Ukrainian hryvnia&UAH\\
East Caribbean dollar&XCD&Nepalese rupee&NPR&United Arab Emirates dirham&AED\\
Egyptian pound&EGP&Netherlands Antillean guilder&ANG&United States dollar&USD\\
Eritrean nakfa&ERN&New Taiwan dollar&TWD&Uruguayan peso&UYU\\
Ethiopian birr&ETB&New Zealand dollar&NZD&Uzbekistani som&UZS\\
Euro&EUR&Nicaraguan cordoba&NIO&Vanuatu vatu&VUV\\
Falkland Islands pound&FKP&Nigerian naira&NGN&Vietnamese dong&VND\\
Fijian dollar&FJD&North Korean won&KPW&Yemeni rial&YER\\
Gambian dalasi&GMD&Norwegian krone&NOK&&\\
Georgian lari&GEL&Omani rial&OMR&&\\
\bottomrule \bottomrule
\end{tabular}
\end{table}

\begin{table}[htbp]
\centering
\caption{Estimated Efficiency Index for gold prices in worldwide currencies}
\label{tab2}
\scriptsize
\begin{tabular}{clc|clc|clc}
\toprule \toprule
Rank&Country&EI&Rank&Country&EI&Rank&Country&EI\\
\midrule \midrule
1&Liberia& 0.1064$\pm$0.0083&49&Albania& 0.2233$\pm$0.0069&97&Dominica& 0.2598$\pm$0.0068\\
2&Seychelles& 0.1167$\pm$0.0067&50&Pakistan& 0.2233$\pm$0.0064&98&Sweden& 0.2601$\pm$0.0054\\
3&Maldives& 0.1447$\pm$0.0060&51&Paraguay& 0.2251$\pm$0.0055&99&Uganda& 0.2606$\pm$0.0056\\
4&Comoros& 0.1473$\pm$0.0076&52&Haiti& 0.2264$\pm$0.0096&100&Barbados& 0.2623$\pm$0.0066\\
5&Somalia& 0.1525$\pm$0.0064&53&Laos& 0.2283$\pm$0.0068&101&Angola& 0.2624$\pm$0.0078\\
6&Tonga& 0.1571$\pm$0.0094&54&Solomon Isl.& 0.2307$\pm$0.0111&102&Denmark& 0.2627$\pm$0.0059\\
7&Mauritania& 0.1572$\pm$0.0063&55&Lebanon& 0.2318$\pm$0.0060&103&Neth. Antilles& 0.2638$\pm$0.0075\\
8&Rwanda& 0.1656$\pm$0.0098&56&Phillipines& 0.2323$\pm$0.0065&104&Saudi Arabia& 0.2639$\pm$0.0081\\
9&Chile& 0.1663$\pm$0.0101&57&Tajikistan& 0.2339$\pm$0.0054&105&Cuba& 0.2640$\pm$0.0064\\
10&S. T. \& Princ& 0.1680$\pm$0.0083&58&Papua N. Guin.& 0.2341$\pm$0.0122&106&Turkey& 0.2659$\pm$0.0055\\
11&Mozambique& 0.1709$\pm$0.0100&59&Mexico& 0.2344$\pm$0.0053&107&India& 0.2665$\pm$0.0053\\
12&Domin. Rep.& 0.1727$\pm$0.0109&60&Bangladesh& 0.2370$\pm$0.0053&108&Croatia& 0.2674$\pm$0.0058\\
13&Libya& 0.1729$\pm$0.0088&61&Peru& 0.2401$\pm$0.0081&109&Tanzania& 0.2684$\pm$0.0075\\
14&Belize& 0.1743$\pm$0.0082&62&Malaysia& 0.2419$\pm$0.0069&110&Lithuania& 0.2687$\pm$0.0056\\
15&Madagascar& 0.1752$\pm$0.0067&63&Bahrain& 0.2436$\pm$0.0064&111&Bulgaria& 0.2696$\pm$0.0049\\
16&Costa Rica& 0.1758$\pm$0.0076&64&Swaziland& 0.2449$\pm$0.0066&112&Azerbaijan& 0.2700$\pm$0.0066\\
17&Fr. Polynesia& 0.1768$\pm$0.0059&65&Taiwan& 0.2450$\pm$0.0099&113&Czech Rep.& 0.2704$\pm$0.0049\\
18&Indonesia& 0.1773$\pm$0.0068&66&Jamaica& 0.2459$\pm$0.0071&114&Canada& 0.2716$\pm$0.0054\\
19&Burundi& 0.1794$\pm$0.0103&67&Eritrea& 0.2461$\pm$0.0069&115&Morocco& 0.2727$\pm$0.0047\\
20&Mauritius& 0.1798$\pm$0.0073&68&Qatar& 0.2461$\pm$0.0068&116&Macau& 0.2732$\pm$0.0062\\
21&Djibouti& 0.1828$\pm$0.0084&69&Trin. \& Tob.& 0.2468$\pm$0.0071&117&Armenia& 0.2733$\pm$0.0095\\
22&Macedonia& 0.1834$\pm$0.0078&70&Argentina& 0.2472$\pm$0.0100&118&Guyana& 0.2742$\pm$0.0074\\
23&Uzbekistan& 0.1836$\pm$0.0075&71&Aruba& 0.2473$\pm$0.0077&119&Serbia& 0.2757$\pm$0.0057\\
24&Iceland& 0.1840$\pm$0.0055&72&Yemen& 0.2476$\pm$0.0070&120&Congo (DRC)& 0.2760$\pm$0.0078\\
25&Cape Verde& 0.1841$\pm$0.0074&73&Ethiopia& 0.2478$\pm$0.0060&121&Singapore& 0.2770$\pm$0.0073\\
26&Sierra Leone& 0.1855$\pm$0.0089&74&Botswana& 0.2479$\pm$0.0092&122&EU& 0.2773$\pm$0.0051\\
27&Ghana& 0.1870$\pm$0.0061&75&Kuwait& 0.2479$\pm$0.0076&123&Norway& 0.2775$\pm$0.0058\\
28&Mongolia& 0.1876$\pm$0.0090&76&Kenya& 0.2481$\pm$0.0058&124&Poland& 0.2781$\pm$0.0069\\
29&Nicaragua& 0.1894$\pm$0.0086&77&Jordan& 0.2490$\pm$0.0065&125&Saint Helena& 0.2818$\pm$0.0111\\
30&Cambodia& 0.1896$\pm$0.0065&78&Thailand& 0.2497$\pm$0.0073&126&Hungary& 0.2819$\pm$0.0055\\
31&Nepal& 0.1905$\pm$0.0055&79&China& 0.2500$\pm$0.0069&127&Nigeria& 0.2824$\pm$0.0092\\
32&Bahamas& 0.1925$\pm$0.0086&80&Namibia& 0.2509$\pm$0.0070&128&Australia& 0.2839$\pm$0.0067\\
33&Brazil& 0.1936$\pm$0.0063&81&Georgia& 0.2513$\pm$0.0078&129&South Korea& 0.2846$\pm$0.0079\\
34&Israel& 0.1950$\pm$0.0068&82&Tunisia& 0.2515$\pm$0.0076&130&Honduras& 0.2904$\pm$0.0082\\
35&Iraq& 0.1987$\pm$0.0078&83&Panama& 0.2528$\pm$0.0069&131&Japan& 0.2905$\pm$0.0077\\
36&Colombia& 0.2010$\pm$0.0076&84&Sri Lanka& 0.2539$\pm$0.0046&132&Switzerland& 0.2921$\pm$0.0056\\
37&Samoa& 0.2042$\pm$0.0091&85&UAE& 0.2551$\pm$0.0072&133&South Africa& 0.2997$\pm$0.0059\\
38&Moldova& 0.2061$\pm$0.0062&86&Turkmenistan& 0.2555$\pm$0.0075&134&New Zealand& 0.3101$\pm$0.0102\\
39&Algeria& 0.2063$\pm$0.0063&87&Lesotho& 0.2555$\pm$0.0065&135&Falkland Isl.& 0.3118$\pm$0.0101\\
40&Ukraine& 0.2103$\pm$0.0105&88&Oman& 0.2559$\pm$0.0061&136&UK& 0.3172$\pm$0.0108\\
41&Cayman Isl.& 0.2105$\pm$0.0059&89&Romania& 0.2560$\pm$0.0070&137&Gibraltar& 0.3177$\pm$0.0098\\
42&Suriname& 0.2120$\pm$0.0090&90&Vietnam& 0.2565$\pm$0.0117&138&Vanuatu& 0.3196$\pm$0.0078\\
43&Fiji& 0.2137$\pm$0.0099&91&Guatemala& 0.2572$\pm$0.0056&139&Belarus& 0.3373$\pm$0.0069\\
44&Kazakhstan& 0.2146$\pm$0.0065&92&Hong Kong& 0.2576$\pm$0.0075&140&Gambia& 0.3530$\pm$0.0095\\
45&Syria& 0.2149$\pm$0.0067&93&Kyrgyzstan& 0.2578$\pm$0.0059&141&Egypt& 0.3574$\pm$0.0116\\
46&Afghanistan& 0.2179$\pm$0.0060&94&Russia& 0.2589$\pm$0.0066&142&Bitcoin& 0.3846$\pm$0.0076\\
47&Brunei& 0.2221$\pm$0.0083&95&North Korean& 0.2596$\pm$0.0064&&\\
48&Uruguay& 0.2226$\pm$0.0056&96&USA& 0.2597$\pm$0.0069&&\\
\bottomrule \bottomrule
\end{tabular}
\end{table}

\end{document}